\documentclass[reqno,11pt]{amsart}
\usepackage{amsfonts, amsthm}

\newtheorem{lemma}{Lemma}
\newtheorem{theorem}{Theorem}

\newtheorem{proposition}{Proposition}
\theoremstyle{remark}
\newtheorem{remark}{Remark}
\numberwithin{equation}{section}

\usepackage{hyperref}

\begin{document}

\title[]{On Darboux non-integrability of the Hietarinta equation}

\author[]{S. Ya. Startsev}

\address{Sergey Ya. Startsev}

\urladdr{\href{http://www.researcherid.com/rid/D-1158-2009}{http://www.researcherid.com/rid/D-1158-2009}}

\urladdr{\href{https://orcid.org/0000-0001-5891-6191}{https://orcid.org/0000-0001-5891-6191}}

\address{\newline\hphantom{iii} Institute of Mathematics,
\newline\hphantom{iii} Ufa Federal Research Centre,
\newline\hphantom{iii} Russian Academy of Sciences}

\begin{abstract}
The autonomous Hietarinta equation is a well-known example of the quad-graph discrete equation which is consistent around the cube. In a recent work, it was conjectured that this equation is Darboux integrable (i.e., for each of two independent discrete variables there exist non-trivial functions that remain unchanged on solutions of the equation after the shift in this discrete variable). We demonstrate that this conjecture is not true for generic values of the equation coefficients.

To do this, we employ two-point invertible transformations introduced by R.I.~Yamilov. We prove that an autonomous difference equation on the quad-graph cannot be Darboux integrable if a transformation of the above type maps solutions of this equation into its solutions again. This implies that the generic Hietarinta equation is not Darboux integrable since the Hietarinta equation in the general case possesses the two-point invertible auto-transformations. Along the way, all Darboux integrable subcases of the Hietarinta equation are found. All of them are reduced by point transformations to already known integrable equations. 

At the end of the article, we also briefly describe another way to prove the Darboux non-integrability of the Hietarinta equation. This alternative way is based on the known fact that a difference substitution relates this equation to a linear one. Thus, the Hietarinta equation gives us an example of a quad-graph equation that is linearizable but not Darboux integrable.
\end{abstract}



\maketitle

{\begin{flushright} {\it Written for the special issue of Ufa Mathematical Journal \\ dedicated to the memory of A.B. Shabat and R.I. Yamilov}\end{flushright}}

\section{Introduction}\label{intr}
The Hietarinta equation 
\begin{equation}\label{hie}
\frac{z_{n,m}+b}{z_{n,m}+a}\, \frac{z_{n+1,m+1}+d}{z_{n+1,m+1}+c} = \frac{z_{n+1,m}+b}{z_{n+1,m}+c}\, \frac{z_{n,m+1}+d}{z_{n,m+1}+a}, \qquad n,m \in \mathbb{Z},
\end{equation}
where the constants $a$, $b$, $c$ and $d$ satisfy the inequality $(a-b)(a-d)(c-b)(c-d) \ne 0$, gives us an important example of the quad-graph difference equation which is consistent around the cube \cite{ABS}. This example was introduced in \cite{hiet} to illustrate that the tetrahedron property is not necessary for the consistency around the cube.

A non-autonomous version of the Hietarinta equation was recently studied in \cite{GS}. It turned out that this non-autonomous version is Darboux integrable. Therefore, at the very end of \cite{GS}, the authors  raise the question of whether the autonomous Hietarinta equation~\eqref{hie} is Darboux integrable too. The main purpose of the present article is to answer this question and to demonstrate that the answer is negative for the generic values of the constants $a$, $b$, $c$ and $d$.

Let us remind that an equation of the form
\begin{equation}\label{hd}
u_{n+1,m+1} = F(u_{n,m}, u_{n+1,m}, u_{n,m+1}), \qquad \frac{\partial F} {\partial u_{n+1,m}}\, \frac{\partial F} {\partial u_{n,m+1}}\, \frac{\partial F} {\partial u_{n,m}} \ne 0
\end{equation}
is called \emph{Darboux integrable} if there exist functions 
$\Omega_{n,m}(u_{n,m},u_{n+1,m},\dots,u_{n+p,m})$, $p>0$, and $\bar{\Omega}_{n,m}(u_{n,m},u_{n,m+1},\dots,u_{n,m+\bar{p}})$, $\bar{p}>0$, such that they essentially depend on $u_{n,m}$ and their last arguments ($u_{n+p,m}$ for $\Omega_{n,m}$ and $u_{n,m+\bar{p}}$ for $\bar{\Omega}_{n,m}$) and satisfy  the relations
\begin{equation}\label{ni}
\Omega_{n,m+1}(u_{n,m+1},u_{n+1,m+1},\dots,u_{n+p,m+1}) = \Omega_{n,m}(u_{n,m},u_{n+1,m},\dots,u_{n+p,m}),
\end{equation}
\begin{equation}\label{mi}
\bar{\Omega}_{n+1,m}(u_{n+1,m},u_{n+1,m+1},\dots,u_{n+1,m+\bar{p}})= \bar{\Omega}_{n,m}(u_{n,m},u_{n,m+1},\dots,u_{n,m+\bar{p}})
\end{equation}
for any $n$, $m$ and any solution $u$ 
of the equation. In other words, $\Omega_{n,m}$ and $\bar{\Omega}_{n,m}$ remain unchanged on solutions of \eqref{hd} after the shifts in $m$ and $n$, respectively. The functions $\Omega_{n,m}$ and $\bar{\Omega}_{n,m}$ are respectively called an \emph{$n$-integral of order $p$} and an \emph{$m$-integral of order $\bar{p}$} for the equation~\eqref{hd}.

It should be explained that we eliminate all variables of the form $u_{n+i,m+j}$, $i\cdot j \ne 0$, via the equation~\eqref{hd} to check whether an relation (in particular, relations~\eqref{ni}, \eqref{mi}) holds on solutions of the equation. I.e., we substitute $F(u_{n,m}, u_{n+1,m}, u_{n,m+1})$ for $u_{n+1,m+1}$,   
\[ F\left(u_{n+1,m}, u_{n+2,m}, F(u_{n,m}, u_{n+1,m}, u_{n,m+1})\right)\quad \text{for}\quad u_{n+2,m+1} \] 
and so on. An relation holds on solutions of the equation iff this relation holds identically after the above substitutions. By the way, these substitutions show why we can assume without loss of generality that the integrals do not depend on the variables of the form $u_{n+i,m+j}$, $i\cdot j \ne 0$. In addition, after eliminating these variables from the defining relations of the integrals, the independence of $\Omega_{n,m}$ and $\bar{\Omega}_{n,m}$ on the variables $u_{n,m+j}$ and $u_{n+i,m}$ directly follows from \eqref{ni} and \eqref{mi}, respectively (see, for example, Lemma~1 in \cite{Stn} if more details is needed).

\begin{remark}
Note that the defining relation~\eqref{ni} can be considered for each $n$ separately, and the corresponding $\Omega_{n,m}$ may be chosen quite differently for different $n$. For example, we can set $\Omega_{n,m}$ equal to constants for some values of $n$ and to non-constant `solutions' of~\eqref{ni} (if they exist) for other $n$. But, since \eqref{hd} does not explicitly depend on $n$, we can choose $\Omega_{n,m}$ the same for all $n$ and assume without loss of generality that it does not explicitly depend on $n$ too. Under this assumption and by using the inequality in \eqref{hd}, it can be proved that $\Omega_{n,m}$ must depend on its first and last arguments for all $n$ and $m$ if $\Omega_{n,m}$ depends on these arguments for at least some $m$ and satisfies~\eqref{ni} for all $m$. Of course, the same (up to the interchange $n \leftrightarrow m$) is also true for \eqref{mi} and $\bar{\Omega}_{n,m}$.
\end{remark}

A simple example of Darboux integrable equation is the `multiplicative' 
discrete wave equation 
\begin{equation}\label{dw}
u_{n+1,m+1}=\frac{u_{n+1,m} u_{n,m+1}}{u_{n,m}}
\end{equation}
admitting the integrals $\Omega_{n,m}= u_{n+1,m}/u_{n,m}$, $\bar{\Omega}_{n,m}= u_{n,m+1}/u_{n,m}$. We can modify \eqref{dw} by the point transformation $\tilde{u}_{n,m} = \zeta_{n,m} u_{n,m}$, where $\zeta_{n,m}=-1$ if both $n$ and $m$ are odd, and $\zeta_{n,m} = 1$ otherwise. This gives us the autonomous equation 
\begin{equation}\label{cdw}
\tilde{u}_{n+1,m+1}=-\frac{\tilde{u}_{n+1,m} \tilde{u}_{n,m+1}}{\tilde{u}_{n,m}}
\end{equation}
that admits the integrals $\Omega_{n,m}= (-1)^m \tilde{u}_{n+1,m}/\tilde{u}_{n,m}$, $\bar{\Omega}_{n,m}= (-1)^n \tilde{u}_{n,m+1}/\tilde{u}_{n,m}$. Note that these integrals are non-autonomous (i.e., they explicitly depend on $m$ and $n$). The last example looks contrived 
(since the equations~\eqref{dw} and \eqref{cdw} are, in fact, the same), but some other autonomous quad-graph equations with non-autonomous integrals can be found, for example, in \cite{GY}. Thus, we may not exclude non-autonomous integrals from consideration when we check an autonomous equation for the Darboux integrability.

Interestingly, the non-autonomous Hietarinta equation in \cite{GS} is reduced to \eqref{cdw} by a non-autonomous M\"obius transformations, while the work~\cite{hiet} relates the equation~\eqref{hie} to \eqref{dw} by a point transformation in the case $|a-c|+|b-d|=0$ only. A less trivial example of Darboux integrable discrete equation is the equation  
\begin{equation}\label{dhi}
\frac{u_{n+1,m+1}}{u_{n,m+1}}=\frac{u_{n+1,m}+1}{u_{n,m}+1}
\end{equation}
admitting the integrals $u_{n,m+1}/(u_{n,m}+1)$ and $(u_{n+2,m}-u_{n+1,m})/(u_{n+1,m}-u_{n,m})$. This equation and its integrals were found in \cite{GY}.
We show below that a point  transformation reduce the equation~\eqref{hie} to \eqref{dhi} in the case $a=c$, $b \ne d$ and to the equation obtained from~\eqref{dhi} by interchanging $n \leftrightarrow m$ in the case $a \ne c$, $b = d$. Hence, the equation~\eqref{hie} is Darboux integrable if $(a-c)(b-d) = 0$. In the present paper, we prove that all other cases of the autonomous Hietarinta equation \eqref{hie} are not Darboux integrable and, moreover, do not admit an integral even in one direction.

As it is demonstrated in \cite{ZhSok,GY_umj}, the smallest orders of the integrals for Darboux integrable equations can be arbitrary high. To prove Darboux non-integrability, we therefore need to make sure that there are no integrals of order $p$ for all $p$. Characteristic algebras \cite{Hab} and Laplace invariants \cite{AdS,stdl} gives us strong and enough constructive necessary conditions for the existence of integrals of order $p$ for the quad-graph equations, but these methods do not solve the problem of the arbitrariness of $p$. This is why the Darboux non-integrability of a quad-graph equation is not always obvious.

To demonstrate the absence of integrals for the generic equation \eqref{hie}, we employ some specific properties of the Heitarinta equation and prove theorems about the absence of integrals for any quad-graph equation with the same properties\footnote{But the author does not know other nonlinear examples of quad-graph equations with these properties.}. These properties are considered in the next section.

\section{Hietarinta equation: transformations and Darboux integrable subcases}\label{hisc}

Following the work \cite{ram}, it is convenient to make the point transformation 
\begin{equation}\label{pth}
z_{n,m}=(d-c u_{n,m})/(u_{n,m}-1) 
\end{equation}
in the equation~\eqref{hie}. (I.e., we denote $(z_{n,m}+d)/(z_{n,m}+c)$ by $u_{n,m}$ and rewrite \eqref{hie} in terms of this new variable.)
In the case $a \ne c$, this gives us the equation
\begin{equation}\label{hiet}
u_{n+1,m+1} (u_{n,m}+B) (u_{n,m+1}+A) = u_{n,m+1} (u_{n+1,m}+B) (u_{n,m}+A),
\end{equation}
where $A=(d-a)/(a-c)$ and $B=(d-b)/(b-c)$. It should be noted that $A \ne B$  because $A-B=(d-c)(b-a)/((a-c)(b-c))$ and the coefficients of~\eqref{hie} satisfy the conditions $d \ne c$ and $b \ne a$. 

If $a=c$, then \eqref{pth} maps the Hietarinta equation~\eqref{hie} into the equation    
\[ \frac{u_{n+1,m+1}}{u_{n,m+1}}=\frac{u_{n+1,m}+B}{u_{n,m}+B}, \]
where the constant $B$ again equals $(d-b)/(b-c)$. The last equation coincides with \eqref{dw} if $B=0$. Otherwise, the scale transformation $u_{n,m} \rightarrow B u_{n,m}$ relates this equation to \eqref{dhi}.

In the case $B=0$, the equation~\eqref{hiet} takes the form 
\[ u_{n+1,m+1} \frac{u_{n,m+1}+A}{u_{n,m+1}} =  u_{n+1,m} \frac{u_{n,m}+A}{u_{n,m}}. \]
Recall that $d \ne a$ for \eqref{hie} and $A \ne 0$ for \eqref{hiet}. Denoting $A/u_{n,m}$ by $v_{n,m}$, we obtain
\begin{equation}\label{dhim}
\frac{v_{n+1,m+1}}{v_{n+1,m}}=\frac{v_{n,m+1}+1}{v_{n,m}+1}. 
\end{equation}
The last equation is Darboux integrable because it coincides with \eqref{dhi} up to the change of notation $v \rightarrow u$ and the interchange $n \leftrightarrow m$.

Summing up the results of three previous paragraphs, we obtain the following statement.
\begin{proposition}\label{p1}
The Hietarinta equation~\eqref{hie} is Darboux integrable if $(a-c)(b-d)=0$. In all other cases, the point transformation~\eqref{pth} reduces \eqref{hie} to the equation~\eqref{hiet} with non-zero constants $A$ and $B$ such that $A \ne B$.
\end{proposition}

Below we prove that the equation~\eqref{hiet} is not Darboux integrable if $A B (A-B) \ne 0$. To do this, we employ two-point invertible transformations. 
Alike transformations for differential-difference and continuous analogues of the quad-graph equations were introduced in the works \cite{Yam90} and \cite{SokSv}, respectively. For the discrete equations~\eqref{hd}, such transformations were, in fact, used in \cite{Yam94} and then considered in a more explicit form in \cite{Sit}. In particular, these transformations for \eqref{hiet} were briefly given in \cite{Sit}, and we reproduce them in the next two paragraphs with detailed calculations (to make it easier for the reader to check them).   

Namely, this  transformation for~\eqref{hiet} is as follows. We can rewrite \eqref{hiet} in the form
\begin{equation}\label{hietm}
u_{n+1,m+1} \frac{u_{n,m+1}+A}{u_{n,m+1}} = \frac{(u_{n+1,m}+B) (u_{n,m}+A)}{u_{n,m}+B}.
\end{equation}
Let us denote
\begin{equation}\label{v0}
v_{n,m} = u_{n+1,m} \frac{u_{n,m}+A}{u_{n,m}} - A.
\end{equation}
Then equations~\eqref{hietm} implies that
\begin{equation}\label{v1}
v_{n,m+1} = \frac{(u_{n+1,m}+B) (u_{n,m}+A)}{u_{n,m}+B} - A
\end{equation}
by virtue (i.e. on solutions) of the equation~\eqref{hiet}. The next step is to solve the system~\eqref{v0}-\eqref{v1} for $u_{n,m}$ and $u_{n+1,m}$ (i.e., to express $u_{n,m}$ and $u_{n+1,m}$ in terms of $v_{n,m}$ and $v_{n,m+1}$). Solving \eqref{v0} for $u_{n+1,m}$, we obtain 
\begin{equation}\label{ev0}
u_{n+1,m} = \frac{u_{n,m}(v_{n,m}+A)}{u_{n,m} + A}.
\end{equation}
The substitution of \eqref{ev0} into~\eqref{v1} results in 
\begin{equation}\label{ev1}
v_{n,m+1} = \frac{u_{n,m}(v_{n,m}+B)}{u_{n,m} + B}.
\end{equation}
Solving~\eqref{ev1} for $u_{n,m}$, we get
\begin{equation}\label{u0}
u_{n,m} = \frac{B v_{n,m+1}}{B+v_{n,m} - v_{n,m+1}}.
\end{equation}
The substitution of \eqref{u0} into~\eqref{ev0} gives us
\begin{equation}\label{u1}
u_{n+1,m} = \frac{B v_{n,m+1} (v_{n,m}+A)}{A(v_{n,m}+B) +(B-A)v_{n,m+1}}.
\end{equation}
If we shift \eqref{u0} by $1$ in $n$ and compare the result with \eqref{u1}, we obtain the equality
\begin{equation}\label{hietn}
\frac{B v_{n+1,m+1}}{B+v_{n+1,m} - v_{n+1,m+1}} = \frac{B v_{n,m+1} (v_{n,m}+A)}{A(v_{n,m}+B) +(B-A)v_{n,m+1}}.
\end{equation}
Solving \eqref{hietn} for $v_{n+1,m+1}$, we see that
\[ v_{n+1,m+1} (v_{n,m}+B) (v_{n,m+1}+A) = v_{n,m+1} (v_{n+1,m}+B) (v_{n,m}+A). \]
The last equation coincides with~\eqref{hiet} up to the change of notation $u \rightarrow v$. Thus, the transformation \eqref{v0} maps solutions of~\eqref{hiet} into solutions of ~\eqref{hiet} again (i.e., \eqref{v0} is a B\"acklund auto-transformation for the Hietarinta equation~\eqref{hiet}).

If we repeat the reasoning of the previous paragraph in the inverse order and then change the notation $u \rightarrow \bar{v}$, $v \rightarrow u$, then we obtain that 
\[ \bar{v}_{n,m} = B u_{n,m+1}/(B+u_{n,m} - u_{n,m+1}) \]
is also an auto-transformation for the equation~\eqref{hiet}. Indeed, starting from~\eqref{hietn} and introducing $u_{n,m}$ by~\eqref{u0}, we obtain \eqref{u1}. Equation~\eqref{ev1} appears by solving~\eqref{u0} for $v_{n,m+1}$, and \eqref{ev0} -- by substituting~\eqref{ev1} into~\eqref{u1}. Eq.~\eqref{ev0} implies~\eqref{v0}, and \eqref{v1} is obtained by the substitution of~\eqref{v0} into~\eqref{ev1}. The comparison of~\eqref{v0} and~\eqref{v1} gives us the Heitarinta equation in the form~\eqref{hietm}.

\begin{remark} Under the condition $B \ne 0$, all formulas~\eqref{hietm}-\eqref{hietn} remain valid even if $A=0$ or $A = B$. Thus, \eqref{v0} and $\bar{v}_{n,m} = B u_{n,m+1}/(B+u_{n,m} - u_{n,m+1})$ are auto-transformations for~\eqref{hiet} in these cases too. But \eqref{hiet} loses the dependence of $u_{n,m+1}$ or $u_{n,m}$ if $A=0$ or $A = B$, respectively.
\end{remark} 

\section{Transformation of integrals}

In this section, we formulate some general propositions that are applicable, in particular, to the Hietarinta equation~\eqref{hiet}. The key propositions (Lemmas~\ref{intf}, \ref{t1}) were proved in \cite{St14} in the case of autonomous integrals. For the reader's convenience and to demonstrate that these propositions remain valid for non-autonomous integrals too, below we reproduce the corresponding (slightly modified) proofs from \cite{St14}.

\begin{lemma}\label{intf} 
Let there exit functions $\varphi(x,y)$, $\varphi_y \ne 0$, and $\psi(x,y)$ such that the right-hand side of~\eqref{hd} satisfies the relation 
\begin{equation}\label{qua}
\varphi(u_{n,m+1}, F(u_{n,m}, u_{n+1,m}, u_{n,m+1}))=\psi(u_{n,m},u_{n+1,m})
\end{equation}
(i.e., $\varphi(u_{n,m+1}, u_{n+1,m+1})=\psi(u_{n,m},u_{n+1,m})$ on solutions of \eqref{hd}). In addition, let the equality $\varphi(x,y)=v$ be uniquely solvable for $y$. Then any $p$-th order $n$-integral for the equation \eqref{hd} can be written in the form
\begin{equation}\label{fff}
\Phi_{n,m}(\varphi(u_{n,m},u_{n+1,m}),\varphi(u_{n+1,m},u_{n+2,m}),\dots,\varphi(u_{n+p-1,m},u_{n+p,m})). 
\end{equation}
\end{lemma}
Note that Lemma~\ref{intf} only defines the form of the integrals (if they exist) but does not guarantee their existence.
\begin{proof}
Solving the equality $\varphi(u_{n,m},u_{n+1,m})=v$ for $u_{n+1,m}$, we obtain 
\begin{equation}\label{uns}
u_{n+1,m}=g(\varphi(u_{n,m},u_{n+1,m}),u_{n,m}).
\end{equation}
Let \eqref{hd} admit a $p$-th order $n$-integral $\Omega_{n,m}$. 
Using expression~\eqref{uns} as well as its consequences derived by shifts in $n$, we rewrite $\Omega_{n,m}$ in terms of $u_{n,m}$, $\varphi(u_{n,m},u_{n+1,m})$ and its shifts in $n$:
\begin{equation}\label{oph}   
\Omega_{n,m} = \Phi_{n,m}(u_{n,m}, \varphi(u_{n,m},u_{n+1,m}),\varphi(u_{n+1,m},u_{n+2,m}),\dots,\varphi(u_{n+p-1,m},u_{n+p,m})).
\end{equation}
The relation~\eqref{qua} implies that $\varphi(u_{n+j,m+1},u_{n+j+1,m+1})=\psi(u_{n+j,m},u_{n+j+1,m})$ on solutions of \eqref{hd}. Therefore, the shift of~\eqref{oph} in $m$ gives us  
\[ \Omega_{n,m+1} = \Phi_{n,m+1}(u_{n,m+1}, \psi(u_{n,m},u_{n+1,m}),\psi(u_{n+1,m},u_{n+2,m}),\dots,\psi(u_{n+p-1,m},u_{n+p,m})). \]
Comparing the last equality and \eqref{oph}, we see that $\Omega_{n,m}$ satisfies the defining relation~\eqref{ni} only if the function $\Phi_{n,m+1}$ in~\eqref{oph} does not depend on its first argument for all $n$ and $m$. 
\end{proof}

Let us consider the equation~\eqref{dhi} as an illustrative example. Subtracting $1$ from both sides of~\eqref{dhi} and then replacing them with their reciprocal values, we rewrite this equation in the form 
\[ \frac{u_{n,m+1}}{u_{n+1,m+1} - u_{n,m+1}} = \frac{u_{n,m} +1}{u_{n+1,m} - u_{n,m}}. \]
Thus, \eqref{dhi} satisfies all conditions of Lemma~\ref{intf} with 
\[ \varphi (u_{n,m},u_{n+1,m}) = \frac{u_{n,m}}{u_{n+1,m} - u_{n,m}}, \qquad \psi (u_{n,m},u_{n+1,m}) = \frac{u_{n,m} +1}{u_{n+1,m} - u_{n,m}}.  \]
(This choice of $\varphi$ and $\psi$ is not most obvious but convenient for the reasoning in the next paragraph.) It is easy to check that 
\begin{equation}\label{exg}
(u_{n+2,m}-u_{n+1,m})/(u_{n+1,m}-u_{n,m}) = (v_{n,m}+1)/v_{n+1,m},
\end{equation}
where $v_{n,m} = \varphi (u_{n,m},u_{n+1,m}) = u_{n,m}/(u_{n+1,m} - u_{n,m})$. Note that the left-hand side of~\eqref{exg} is an $n$-integral of \eqref{dhi} and the right-hand side is the representation in the form~\eqref{fff} for this integral.

Applying the scheme of the invertible two-point transformations from Section~\ref{hisc}, we see that $v_{n,m} = u_{n,m}/(u_{n+1,m} - u_{n,m})$ maps solutions of~\eqref{dhi} into solutions of the equation~\eqref{dhim} and the right-hand side of~\eqref{exg} is an $n$-integral for~\eqref{dhim}.
The latter is a particular case of a more general fact formulated in the following Lemma.
\begin{lemma}\label{t1}
Let equation~\eqref{hd} satisfy the assumptions of Lemma~\ref{intf}, the functions $\varphi(x,y)$, $\psi(x,y)$ be functionally independent and the transformation $v_{n,m}=\varphi(u_{n,m},u_{n+1,m})$ map solutions of~\eqref{hd} into solutions of an equation
\begin{equation}\label{pqd}
v_{n+1,m+1}=Q(v_{n,m},v_{n+1,n},v_{n,m+1}).
\end{equation}
Then the equation~\eqref{hd} admits an $n$-integral of order $p>1$ only if the equation~\eqref{pqd} admits an $n$-integral of order $p-1$.  
\end{lemma}
\begin{proof}
Let $\Omega_{n,m}$ be a $p$-th order $n$-integral of~\eqref{hd}. Lemma~\ref{intf} implies that
\begin{equation}\label{oph1} 
\Omega_{n,m}= \Phi_{n,m}(v_{n,m},v_{n+1,m},\dots,v_{n+p-1,m}),
\end{equation}
 where $v_{i,j}=\varphi(u_{i,j},u_{i+1,j})$. Let $T_n^{k}$ denote the combination of the shift by $k$ in $n$  and the elimination of the variables $v_{n+i,m+1}$, $i>0$, by using the equation~\eqref{pqd}. Since $v_{n,m}$ satisfies \eqref{pqd} for any solution of~\eqref{hd}, we have
\[ v_{n+1,m+1} = T_{n}^1 (v_{n,m+1}) = Q(v_{n,m}, v_{n+1,m}, v_{n,m+1}), \]
\[ v_{n+2,m+1} = T_{n}^1 (Q) = Q\left(v_{n+1,m}, v_{n+2,m}, Q(v_{n,m}, v_{n+1,m}, v_{n,m+1})\right), \]
\[ v_{n+k+1,m+1} = T_{n}^k (Q) = Q(v_{n+k,m}, v_{n+k+1,m}, T_{n}^{k-1}(Q)), \qquad k >1. \]
Substituting these formulas into the defining relation for the integral~\eqref{oph1}, we obtain
\begin{equation}\label{defe}
\Phi_{n,m+1} (v_{n,m+1},Q,\dots,T_n^{p-2}(Q))=\Phi_{n,m}(v_{n,m},v_{n+1,m},\dots,v_{n+p-1,m}).
\end{equation}
This equality holds when $v_{n,m+1}=\varphi(u_{n,m+1},u_{n+1,m+1})=\psi(u_{n,m},u_{n+1,m})$ (see~\eqref{qua}) and $v_{n+\ell,m}=\varphi(u_{n+\ell,m},u_{n+\ell+1,m})$,  $\ell=\overline{0,p-1}$ (other variables $v_{i,j}$ are absent in~\eqref{defe}). But $\psi(u_{n,m},u_{n+1,m})$ and $\varphi(u_{n+\ell,m},u_{n+\ell+1,m})$ are functionally independent and, hence, \eqref{defe} must hold identically for arbitrary $v_{n,m+1}$ and $v_{n,m+\ell}$. Thus, $\Phi_{n,m}(v_{n,m},v_{n+1,m},\dots,v_{n+p-1,m})$ is an $n$-integral of the equation~\eqref{pqd}. Taking \eqref{oph1} into account, we see that $\Phi_{n,m}$ essentially depends on $v_{n,m}$ and $v_{n+p-1,m}$ if $\partial \Omega_{n,m} / \partial u_{n,m} \ne 0$ and $\partial \Omega_{n,m} / \partial u_{n+p,m} \ne 0$. The integral $\Phi_{n,m}$ therefore has order $p-1$.
\end{proof}

\begin{remark} In contrast to~\eqref{hd}, we do not assume 
\[ \frac{\partial Q} {\partial v_{n+1,m}}\, \frac{\partial Q} {\partial v_{n,m+1}}\, \frac{\partial Q} {\partial v_{n,m}} \ne 0 \] 
for~\eqref{pqd}. But this inequality follows from assumptions of Lemma~\ref{t1}. Indeed, substituting 
\[ v_{n,m} = \varphi (u_{n,m}, u_{n+1,m}), \quad v_{n,m+1} = \psi (u_{n,m}, u_{n+1,m}) \]
into \eqref{pqd}, we obtain
\begin{equation}\label{rem3}
\psi(u_{n+1,m}, u_{n+2,m}) = Q \left(\varphi (u_{n,m}, u_{n+1,m}),\varphi (u_{n+1,m}, u_{n+2,m}), \psi (u_{n,m}, u_{n+1,m})\right)
\end{equation}
and see that the last relation holds only if $Q$ essentially depends on its second argument. The functional independence of $\varphi (u_{n+1,m}, u_{n+2,m})$ and $\psi (u_{n+1,m}, u_{n+2,m})$ excludes the case when $Q$ does not depend on both first and third arguments. Differentiating relation~\eqref{rem3} with respect to $u_{n,m}$, we see that this relation cannot hold if $Q$ does not depends of its first or third arguments (since $\varphi (u_{n,m}, u_{n+1,m})$, $\psi (u_{n,m}, u_{n+1,m})$, $\varphi (u_{n+1,m}, u_{n+2,m})$ are functionally independent and these functions must essentially depend on their first argument to be compatible with both \eqref{qua} and the inequality in \eqref{hd}). 
\end{remark}

The next proposition is a direct consequence of Lemma~\ref{t1}.
\begin{theorem}\label{te1}
Let equation~\eqref{hd} satisfy the assumptions of Lemma~\ref{intf}, the functions $\varphi(x,y)$, $\psi(x,y)$ be functionally independent and the transformation $v_{n,m}=\varphi(u_{n,m},u_{n+1,m})$ map solutions of~\eqref{hd} into solutions of \eqref{hd} again. Then the equation~\eqref{hd} does not admit $n$-integrals.   
\end{theorem}
Roughly speaking, the above theorem means that a quad-graph equation has no $n$-integrals if it admits a two-point invertible (in the sense of \cite{SokSv,Sit}) auto-transformation depending on the shift of $u_{n,m}$ in $n$.
\begin{proof} Assume the contrary. Let \eqref{hd} have an $n$-integrals of order $p$. If $p>1$, then, applying Lemma~\ref{t1} several times, we obtain that \eqref{hd} admits a first-order $n$-integral.

Any first-order integral of~\eqref{hd} has the form $\Phi_{n,m}(\varphi(u_{n,m},u_{n+1,m}))$ by Lemma~\ref{intf}, and \eqref{qua} implies that the defining relation for this integral takes the form 
\[ 
\Phi_{n,m+1}(\psi(u_{n,m},u_{n+1,m}))=\Phi_{n,m}(\varphi(u_{n,m},u_{n+1,m})). \]
But this contradicts the functional independence of $\psi$ and $\varphi$.
\end{proof}

Making the interchange $n \leftrightarrow m$ in Lemmas~\ref{intf},\ref{t1}, Theorem~\ref{te1} and in their proofs, we also prove the following analogue of Theorem~\ref{te1} for $m$-integrals.

\begin{theorem}\label{te1m}  
Let there exit functionally independent functions $\bar{\varphi}(x,y)$ and $\bar{\psi}(x,y)$ such that the right-hand side of~\eqref{hd} satisfies the relation 
\[ \bar{\varphi}(u_{n+1,m}, F(u_{n,m}, u_{n+1,m}, u_{n,m+1}))=\bar{\psi}(u_{n,m},u_{n,m+1}) \]
(i.e. $\bar{\varphi}(u_{n+1,m}, u_{n+1,m+1})=\bar{\psi}(u_{n,m},u_{n,m+1})$ on solutions of \eqref{hd}). In addition, let the equality $\bar{\varphi}(x,y)=\bar{v}$ be uniquely solvable for $y$. Then the equation~\eqref{hd} does not admit $m$-integrals if the transformation $\bar{v}_{n,m}=\bar{\varphi}(u_{n,m},u_{n,m+1})$ maps solutions of~\eqref{hd} into solutions of \eqref{hd} again. 
\end{theorem}

As it is demonstrated in Section~\ref{hisc}, the equation~\eqref{hiet} satisfies all assumptions of Theorems~\ref{te1}, \ref{te1m} in the case $A B (A-B) \ne 0$. The corresponding functions $\varphi$, $\psi$, $\bar{\varphi}$ and $\bar{\psi}$ are 
\[ \varphi(u_{n,m},u_{n+1,m}) =u_{n+1,m} \frac{u_{n,m}+A}{u_{n,m}} - A, \quad
\psi(u_{n,m},u_{n+1,m}) = \frac{(u_{n+1,m}+B) (u_{n,m}+A)}{u_{n,m}+B} - A, \]
\[ \bar{\varphi}(u_{n,m},u_{n,m+1}) =\frac{B u_{n,m+1}}{B+u_{n,m} - u_{n,m+1}}, \quad \bar{\psi}(u_{n,m},u_{n,m+1}) = \frac{B u_{n,m+1} (u_{n,m}+A)}{A(u_{n,m}+B) +(B-A)u_{n,m+1}}. \]
Thus, taking Proposition~\ref{p1} into account, we obtain the following conclusion.
\begin{proposition}
The Hietarinta equation~\eqref{hie} has no $n$- and $m$-integrals if $(a-c)(b-d) \ne 0$.
\end{proposition}

Theorems~\ref{te1} and \ref{te1m} are also applicable to the linear equation
\begin{equation}\label{linex}
u_{n+1,m+1} = \alpha u_{n+1,m} + \beta u_{n,m+1} + \gamma u_{n,m} 
\end{equation}
with constant coefficients $\alpha$, $\beta$ and $\gamma$. Indeed, 
\[ \varphi(u_{n,m},u_{n+1,m}) =u_{n+1,m} - \beta u_{n,m}, \quad
\psi(u_{n,m},u_{n+1,m}) = \alpha u_{n+1,m} + \gamma u_{n,m},  \]
\[ \bar{\varphi}(u_{n,m},u_{n,m+1}) = u_{n,m+1} - \alpha u_{n,m}, \quad \bar{\psi}(u_{n,m},u_{n,m+1}) =  \beta u_{n,m+1} + \gamma u_{n,m} \]
for this equation. The transformations  $v_{n,m} =u_{n+1,m} - \beta u_{n,m}$, $\bar{v}_{n,m} =u_{n,m+1} - \alpha u_{n,m}$, which coincide with the discrete Laplace transformations (see \cite{Nov,AdS}), map solutions of~\eqref{linex} into solutions of~\eqref{linex} again. Thus, Theorems~\ref{te1} and \ref{te1m} imply that \eqref{linex} has no integrals in the case $\gamma + \alpha \beta \ne 0$ because both the pairs $\varphi$,~$\psi$ and $\bar{\varphi}$,~$\bar{\psi}$ are functionally independent under this condition. 

\begin{remark}\label{r4}
As it was demonstrated in \cite{AdS,stdl}, vanishing a Laplace invariant is a necessary condition for the existence of autonomous integrals of~\eqref{hd}. Checking the proof of Proposition~2 in \cite{stdl}, we can make sure that this necessary condition remains valid for non-autonomous integrals of autonomous equations. Since all the Laplace invariants of~\eqref{linex} are equal to $\gamma + \alpha \beta$, we see that the Laplace invariants provide us with an alternative way to prove the absence of integrals for the equation~\eqref{linex} in the case $\gamma + \alpha \beta \ne 0$. And this gives us another way to prove the Darboux non-integrability of the generic Hietarinta equation if we employ the fact \cite{ram} that the transformation $u_{n,m}=v_{n+1,m}/v_{n,m} - A$ maps solutions of the linear equation 
\begin{equation}\label{lineh}
 v_{n+1,m+1} = v_{n+1,m} + A v_{n,m+1} + (B-A) v_{n,m}   
\end{equation} into solutions of \eqref{hiet}. Indeed, since all the Laplace invariants of~\eqref{lineh} are equal to $B$, this linear equation is not Darboux integrable in the case $B \ne 0$ and the Darboux integrability of the corresponding Hietarinta equation~\eqref{hiet} contradicts the following obvious statement (cf. Lemma~1 in \cite{GY_umj}).
\end{remark}

\begin{proposition}\label{fin}
Let a transformation 
\begin{equation}\label{tra}
u_{n,m}=\phi(v_{n,m},v_{n+1,m}, \dots, v_{n+k,m}), \qquad k>0, \quad  \frac{\partial \phi}{\partial v_{n,m}} \frac{\partial \phi}{\partial v_{n+k,m}} \ne 0,
\end{equation}
map solutions of an equation
\begin{equation}\label{q2}
v_{n+1,m+1} = Q (v_{n,m}, v_{n+1,m}, u_{n,m+1}) 
\end{equation}
into solutions of \eqref{hd} and let the equation~\eqref{hd} admit an $n$-integral of order $p$. Then the equation~\eqref{q2} possesses an $n$-integral of order $p+k$.
\end{proposition}
\begin{proof}
Let $\Omega_{n,m}$ be an $n$-integral for \eqref{hd}. Then the defining relation~\eqref{ni} holds for any solution of~\eqref{hd} and, in particular, for all solutions obtained by formula~\eqref{tra} from solutions of~\eqref{q2}. Therefore,
\[ \Omega_{n,m}\left(\phi(v_{n,m},v_{n+1,m}, \dots, v_{n+k,m}), \dots, \phi(v_{n+p,m},v_{n+p+1,m}, \dots, v_{n+p+k,m}) \right) \]
is an $n$-integral of~\eqref{q2}.
\end{proof}

Thus, the Hietarinta equation gives us an example of a quad-graph equation that is linearizable but not Darboux integrable.

It should be noted that Remark~\ref{r4} describes a sketch for the proof of the Darboux non-integrability of \eqref{hiet} in a way very similar to that was used in \cite{GY_umj} for estimating the minimal orders of integrals for a particular linearizable quad-graph equation. In the present paper, the author prefers another way that does not require introducing the Laplace invariants and therefore seems more self-contained.

\bigskip

\end{document}